\documentclass[12pt]{article}
\usepackage{epsf}
\usepackage{graphicx}
\usepackage{subfigure}
\usepackage{feynmf}
\usepackage{tikz}
\usepackage{epstopdf}
\usepackage{amsmath}
\usepackage{float}

\newcommand\pubnumber{WSU--HEP--XXYY}
\newcommand\pubdate{\today}

\def\pitt{Pittsburgh Particle Physics Astrophysics and Cosmology Center (PITT PACC),\\
Department of Physics and Astronomy,\\ University of Pittsburgh, Pittsburgh, PA 15260, USA \vspace*{.2cm}}

\def\Title#1{\begin{center} {\Large #1 } \end{center}}
\def\Author#1{\begin{center}{ \sc #1} \end{center}}
\def\Address#1{\begin{center}{ \it #1} \end{center}}

\newcommand\pubblock{\rightline{\begin{tabular}{l} \pubnumber\\
         \pubdate  \end{tabular}}}
\newenvironment{Abstract}{\begin{quotation}  }{\end{quotation}}
\newenvironment{Presented}{\begin{quotation} \begin{center} 
             PRESENTED AT\end{center}\bigskip 
      \begin{center}\begin{large}}{\end{large}\end{center} \end{quotation}}
\def\Acknowledgements{\bigskip  \bigskip \begin{center} \begin{large}
             \bf ACKNOWLEDGEMENTS \end{large}\end{center}}

\newcommand{\beq}{\begin{equation}}
\newcommand{\eeq}{\end{equation}}

\def\beq{\begin{equation}}
\def\eeq#1{\label{#1}\end{equation}}
\def\eeqn{\end{equation}}

\def\beqa{\begin{eqnarray}}
\def\eeqa#1{\label{#1}\end{eqnarray}}
\def\eeqan{\end{eqnarray}}

\let\bar=\overbar

\def\Dslash{\not{\hbox{\kern-4pt $D$}}}
\def\dslash{\not{\hbox{\kern-2pt $\del$}}}

\def\msb{{\bar{\ssstyle M \kern -1pt S}}}

\begin{document}
\begin{titlepage}

\pubblock

\vfill

\Title{Production asymmetries of $D^{\pm}$, $\Lambda_c^{+}/\Lambda_c^{-}$ and $\Lambda_b^0/\overline{\Lambda}_b^0$ at the LHC from
heavy quark recombination mechanism}
\vfill
\Author{W. K. Lai}
\Address{\pitt}
\vfill
\begin{Abstract}

The asymmetry in the forward region production cross section of $D^{\pm}$ 
is calculated using the heavy quark recombination mechanism for $pp$ collisions at $7$~TeV. 
By suitable choices of four nonperturbative parameters, our calculated results can reproduce
those obtained at LHCb.
We find $A_p\sim-1\%$ when integrated over $2.0\textrm{ GeV}<p_T<18\textrm{ GeV}$ and $2.2<\eta<4.75$,
which agrees with $A_p=-0.96\pm0.26\pm0.18\%$ as measured by LHCb. Furthermore, the calculated distributions
in $\eta$ and $p_T$ agree reasonably well with those obtained at LHCb. Using the heavy quark recombination mechanism, we also 
make predictions on the production asymmetries of $\Lambda_c^{+}/\Lambda_c^{-}$ and $\Lambda_b^0/\overline{\Lambda}_b^0$ for 
$pp$ collisions at $7$~TeV and $14$~TeV in the forward region. We find that the integrated asymmetries for these $\Lambda$ baryons 
in the LHCb region are of the order of $\sim1-3\%$ and should be measurable. 

\end{Abstract}
\vfill
\begin{Presented}
The 7th International Workshop on Charm Physics (CHARM 2015)\\
Detroit, MI, 18-22 May, 2015
\end{Presented}
\vfill
\end{titlepage}
\def\thefootnote{\fnsymbol{footnote}}
\setcounter{footnote}{0}
%

\section{Introduction}
One of the simplest signals for CP violation in charm is obtained by comparing partial decay widths of charm mesons to those of anticharm mesons. While CPT symmetry requires the total widths of $D$ and $\overline D$ to be the same, the partial decay widths $\Gamma(D \to f)$ and $\Gamma(\overline D \to \overline f)$ are different in the presence of CP violation, which is signaled by a nonzero value of the asymmetry
\begin{equation}
a^f_{CP} = \frac{\Gamma(D \to f)-\Gamma(\overline D \to \overline f)}{\Gamma(D \to f)+\Gamma(\overline D \to \overline f)}\,.
\end{equation}
This signal is reasonably robust for $D^+/D^-$ mesons, provided that the number of decaying particles and antiparticles is the same. However, at the Large Hadron Collider (LHC), the number of produced $D^+$ and $D^-$ mesons might not be the same due to the fact that the initial state contains two protons. With CP-violating asymmetries expected to be at the per mille levels~\cite{Artuso:2008vf}, it is important to examine the production asymmetry of $D$ mesons both experimentally and theoretically.

Indeed, fixed-target experiments have already observed large asymmetries of charmed mesons and baryons in the forward region. In hadroproduction, the charmed hadrons are preferentially produced with a light valence quark of the same type as what appears in the hadronic beam, for example \cite{Aitala:1996hf}. This has been termed the ``leading particle effect". More recently, a similar asymmetry in $D^\pm$ production, defined as
\begin{equation}
A_p=\frac{\sigma(D^+)-\sigma(D^-)}{\sigma(D^+)+\sigma(D^-)}\,,  \label{eq:asymmetry}
\end{equation}
has been measured in the forward region to be $\sim -1\%$ by the LHCb Collaboration~\cite{LHCb:2012fb}. In Section~\ref{D_asymmetry}, we report our findings of explaining this $D^\pm$ production asymmetry at LHCb by the heavy quark recombination mechanism~\cite{D_asym}. In Section~\ref{Lambda_asymmetry}, using the heavy quark recombination mechanism, we predict the production asymmetries of $\Lambda_c^{+}/\Lambda_c^{-}$ and $\Lambda_b^0/\overline{\Lambda}_b^0$ for $pp$ collisions at $7$~TeV and $14$~TeV in the forward region~\cite{Lambda_asym}. We conclude in Section~\ref{conclusion}.

\section{$D^\pm$ production asymmetry from heavy quark recombination mechanism}\label{D_asymmetry}
Factorization theorems of perturbative QCD~\cite{Collins:1985gm} state that heavy hadron production cross section can be written in a factorized form. At the LHC, the cross section for producing a $D$ ($c\bar{q}$) meson in a $pp$ collision, at leading order in a $1/p_T$ expansion, is given by
\begin{equation}
d\sigma[pp\rightarrow D+X]=\sum\limits_{i,j}f_{i/p}\otimes f_{j/p}\otimes d\hat{\sigma}[ij
\rightarrow c+X]\otimes D_{c\rightarrow D}\,,    \label{eq:pQCD}
\end{equation}
where $f_{i/p}$ is the parton distribution function for parton $i$ in the proton, $d\hat{\sigma}
(ij\rightarrow c+X)$ is the partonic cross section and $D_{c\rightarrow D}$ is the 
fragmentation function describing hadronization of a $c$ quark into a $D$ meson. The corresponding equation for 
$\overline{D}$ is obtained by replacing $c$ by $\bar{c}$ and $D$ by $\overline{D}$. Owing to charge conjugation symmetry and that $f_{c/p}=f_{\bar{c}/p}$, perturbative QCD Eq.~(\ref{eq:pQCD}) predicts 
that $A_p=0$, which is at least true at leading order in the $1/p_T$ expansion.

\begin{figure}
\begin{center}
\subfigure[]{
\includegraphics[scale=0.21]{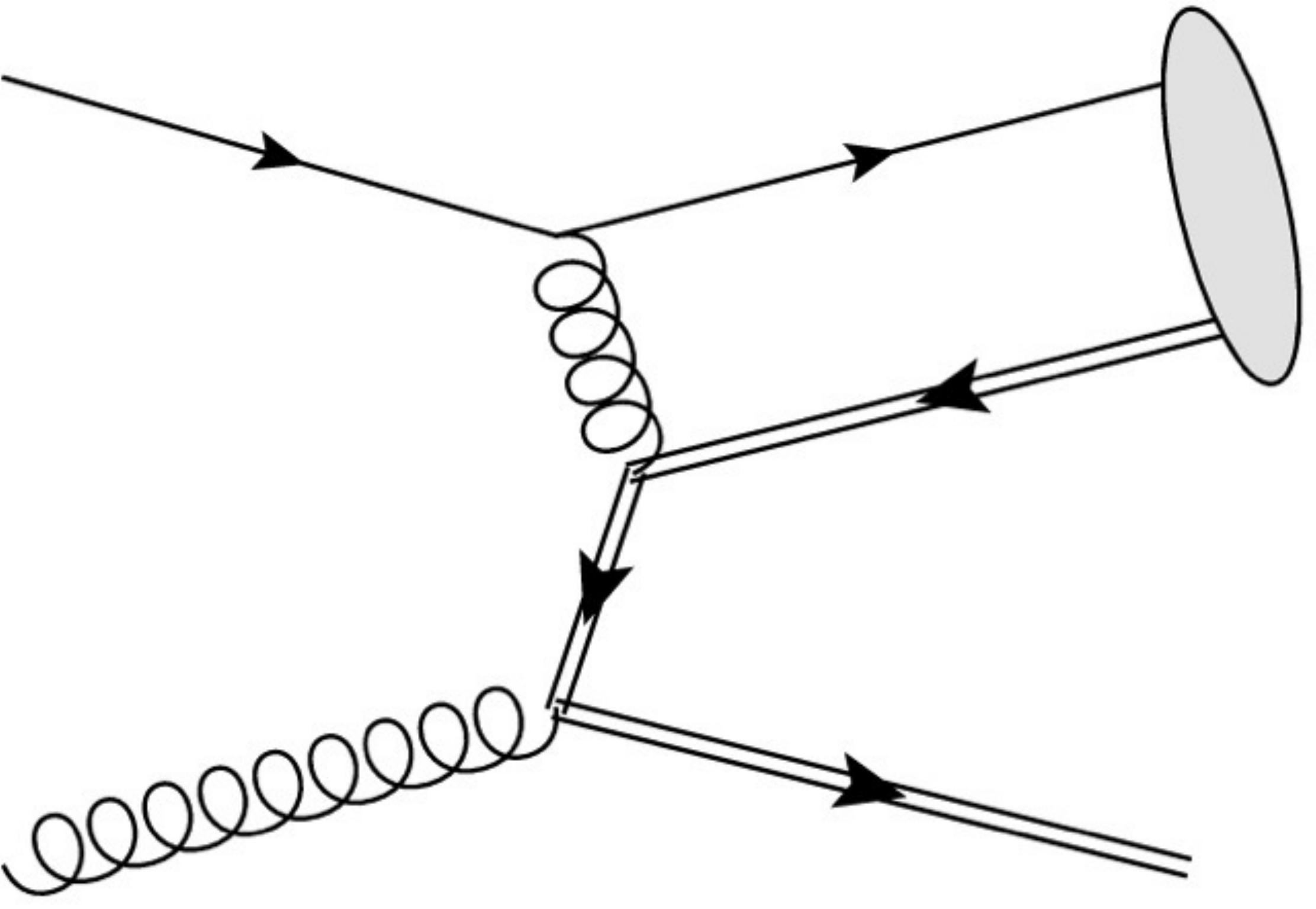}}\label{fg:recomb_qg}
\hskip .5in
\subfigure[]{
\includegraphics[scale=0.04]{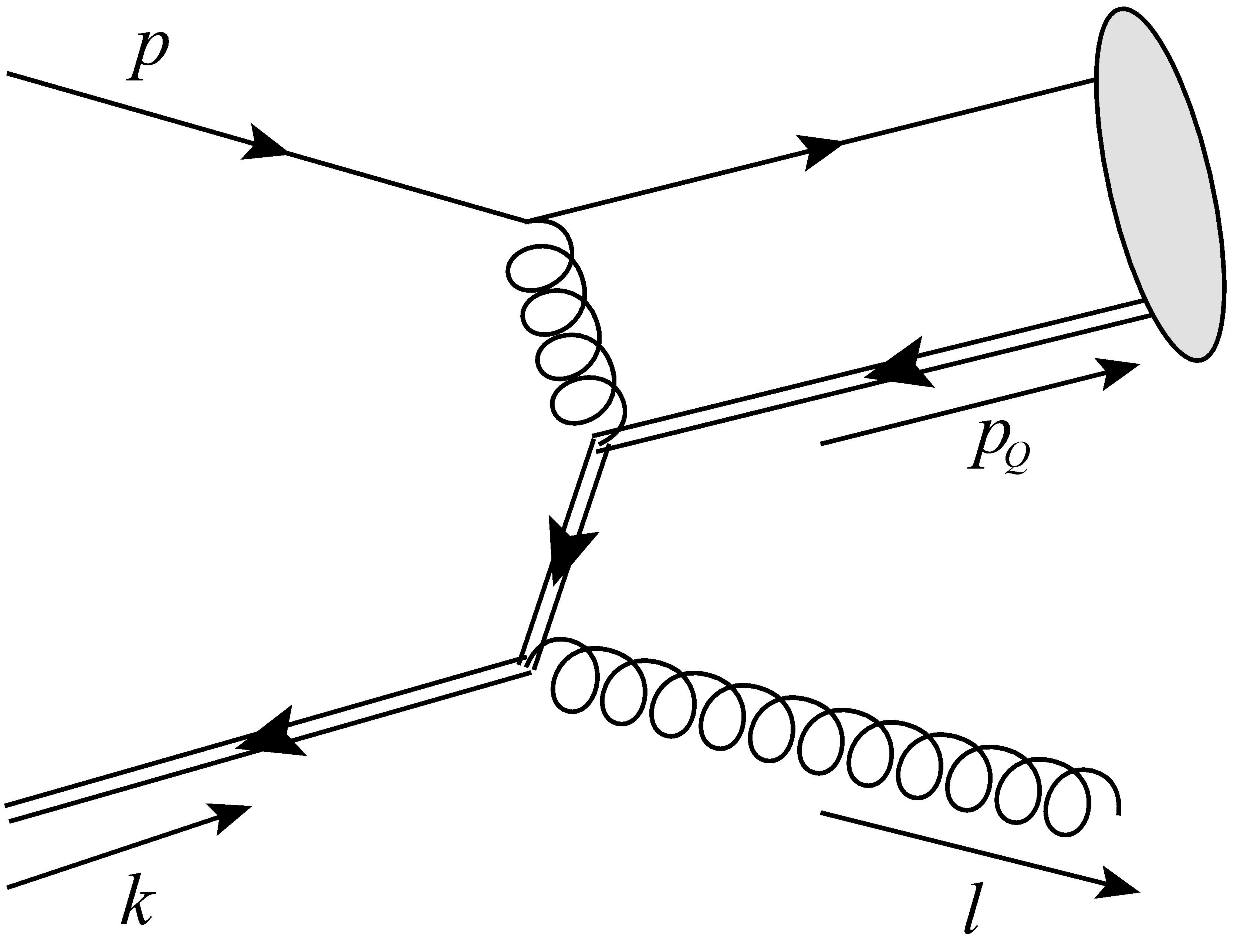}}\label{fg:recomb_Qq}
\vskip .05in  
\caption[]{Diagrams for production of a $\overline{D}$ meson by the heavy quark recombination mechanism for (a) $qg\rightarrow (\bar c q)^n+c$
and (b) $q\bar{c}\rightarrow (\bar c q)^n+g$. Each process has five diagrams. Single lines represent light quarks, 
double lines heavy quarks, and the shaded blob the $\overline{D}$ meson. }
\label{fg:recomb}
\end{center}
\end{figure}

To reconcile the experimental observations with QCD, we note that there are corrections to Eq.~(\ref{eq:pQCD}) that scale as powers of 
$\Lambda_{\rm QCD}/m_c$ and $\Lambda_{\rm QCD}/p_T$. In principle, one can expect nonvanishing power-suppressed contributions to 
$A_p$ at low $p_T$.  A QCD-based model for these power corrections is the heavy quark recombination 
mechanism~\cite{Braaten:2001bf,Braaten:2001uu,Braaten:2002yt,Braaten:2003vy}.  In this scenario, a light quark involved in the hard 
scattering process combines with the heavy quark produced in that interaction to form the final state meson, leading to corrections of order 
$\Lambda_{\rm QCD} m_c/p_T^2$. This contribution to the cross section is given by (Fig.~\ref{fg:recomb}~(a))
\begin{equation}
d\hat\sigma[\overline D]=d\hat{\sigma}[qg\rightarrow (\bar c q)^n+c]\rho[(\bar cq)^n\rightarrow \overline D]\,, \label{eq:recomb}
\end{equation}
where $(\bar cq)^n$ indicates that the light quark of flavor $q$ with momentum of order $\Lambda_{\rm QCD}$ in the $\bar c$ rest frame is produced in the state $n$, where $n$ labels the color and angular momentum quantum numbers of the quark pair. The cross section is factored into a 
perturbatively calculable piece $d\hat{\sigma}[qg\rightarrow (\bar c q)^n+c]$ and a nonperturbative factor $\rho[(\bar cq)^n\rightarrow \overline D]$ encoding the probability for the quark pair with quantum number $n$ to hadronize into a final state including a $\overline D$.  
The perturbative piece was calculated to lowest order in~\cite{Braaten:2001bf}.
Equation~(\ref{eq:recomb}) must then be convoluted with the proton parton distribution functions to get the final hadronic cross section.
Besides the $qg\rightarrow (\bar c q)^n+c$ process, there are also contributions from $q\bar{c}\rightarrow (\bar c q)^n+g$, as shown in Fig.~\ref{fg:recomb}~(b). Using the method introduced in \cite{Braaten:2001bf}, we calculate the partonic cross sections from initial state charm~\cite{D_asym}.

The $c$ quark in Eq.~(\ref{eq:recomb}) could fragment into a $D$ meson. Thus, to get the full rate due to recombination for producing $\overline D$ mesons, we also need to account for the contribution where a light antiquark comes from the proton, while the $\bar c$ fragments into a $\overline{D}$. We thus have three contributions:
\begin{subequations}
\begin{align}
(a)\phantom{aaaa}d\hat\sigma[\overline D]&=d\hat{\sigma}[qg\rightarrow (\bar c q)^n+c]\rho[(\bar c q)^n\rightarrow \overline D]\,,\label{eq:recombination} \\
(b)\phantom{aaaa}d\hat\sigma[\overline D]&=d\hat{\sigma}[q\bar{c}\rightarrow (\bar c q)^n+g]\rho[(\bar c q)^n\rightarrow \overline D]\,,
\label{eq:recombination_Qq} \\
(c)\phantom{aaaa}d\hat\sigma[\overline D]&=d\hat{\sigma}[\bar qg\rightarrow (c\bar q)^n+\bar{c}]\rho[(c\bar q)^n\rightarrow H]\otimes
D_{\bar c\rightarrow \overline D}\,,
\label{eq:recombinationfragment}
\end{align}
\end{subequations}
where $H$ can be any hadron. The recombination cross section for producing a $D$ is obtained by taking the charge conjugate of the above equations.
Below, we will neglect $C$ violation and take $\rho[(\bar{c}q)^n\rightarrow \overline{D}]=\rho[(c\bar{q})^n\rightarrow D]$. For simplicity,
in process (c) we will restrict $H$ to be $D$ or $D^*$ only and sum over $\bar{q}=\bar{u},\bar{d}$ and $\bar{s}$ with $SU(3)$ flavor symmetry assumed.   

The leading contributions to productions of $D^\pm$ mesons by heavy quark recombination consists of four possible options of $n$:
\begin{align}
\rho^{sm}_1&=\rho[c\bar{d}(^1S_0^{(1)})\rightarrow D^+]\,,
&\rho^{sf}_1&=\rho[c\bar{d}(^3S_1^{(1)})\rightarrow D^+]\,, \nonumber \\
\rho^{sm}_8&=\rho[c\bar{d}(^1S_0^{(8)})\rightarrow D^+]\,,  
&\rho^{sf}_8&=\rho[c\bar{d}(^3S_1^{(8)})\rightarrow D^+]\,. \label{eq:rhos}
\end{align}
These nonperturbative parameters must be extracted from data.   
Neglecting $\rho^{sf}_1$ and $\rho^{sf}_8$, the combination $\rho^{sm}_1+\rho^{sm}_8/8$ was determined to be $0.15$ by fitting to the 
E687 and E691 fixed-target photoproduction data~\cite{Braaten:2001uu}. Neglecting $\rho^{sm}_8$, $\rho^{sf}_1$ and $\rho^{sf}_8$, the parameter
$\rho^{sm}_1$ was determined to be $0.06$ by fitting to data from the E791 experiment~\cite{Braaten:2002yt}. 
In this paper, we take $\rho^{sm}_1\sim0.06$
and $\rho^{sm}_8\sim0.7$. It turns out that these two contributions only account for $\sim30\%$ of the measured asymmetry 
$A_p=(-0.96\pm 0.26\pm0.18)\%$ at LHCb in Ref.~\cite{LHCb:2012fb}. Therefore, we include $\rho^{sf}_1$ and $\rho^{sf}_8$ and choose values of similar size as the spin-matched parameters. 
We also include feed down from $D^{*\pm}$. From heavy quark spin symmetry, we have
\begin{align}
\rho[c\bar{d}(^1S_0^{(c)})\rightarrow D^+]&=\rho[c\bar{d}(^3S_1^{(c)})\rightarrow D^{*+}]\,, \nonumber\\
\rho[c\bar{d}(^3S_1^{(c)})\rightarrow D^+]&=\rho[c\bar{d}(^1S_0^{(c)})\rightarrow D^{*+}]\,. \label{eq:heavy_quark_sym}
\end{align}
We use MSTW 2008 LO central PDFs with $m_c = 1.275$ GeV and the Peterson parametrization for the fragmentation function is used for $D_{c\rightarrow H}$:
\begin{equation}
D_{c\rightarrow H}(z)=\frac{N_H}{z\left(1-\frac{1}{z}-\frac{\epsilon_c}{1-z}\right)^2}\,.  \label{eq:Peterson}
\end{equation}
$\epsilon_c\sim (m_q/m_c)^2$ was measured to be $0.062\pm0.007$ for the $D^{*+}$ meson~\cite{Chekanov:2008ur}. Charge conjugation symmetry and approximate heavy quark symmetry implies that $\epsilon_c$ is approximately the same for $D^{\pm}$ and $D^{*\pm}$. We will take $\epsilon_c=0.06$ for both $D^{\pm}$ and $D^{*\pm}$. $N_H$ are determined by the averages of the measured fragmentation probabilities listed in~\cite{Abramowicz:2013eja}. 
For the perturbative QCD rate, Eq.~(\ref{eq:pQCD}), which has no asymmetry if we ignore $C$ violation but enters into the denominator of     
Eq.~(\ref{eq:asymmetry}), we use the LO cross section. The factorization scale is set to be $\mu_f=\sqrt{p_T^2+m_c^2}$.  

\begin{figure}[t]
\begin{center}
\subfigure[]{
      \includegraphics[width=0.45\textwidth,angle=0,clip]{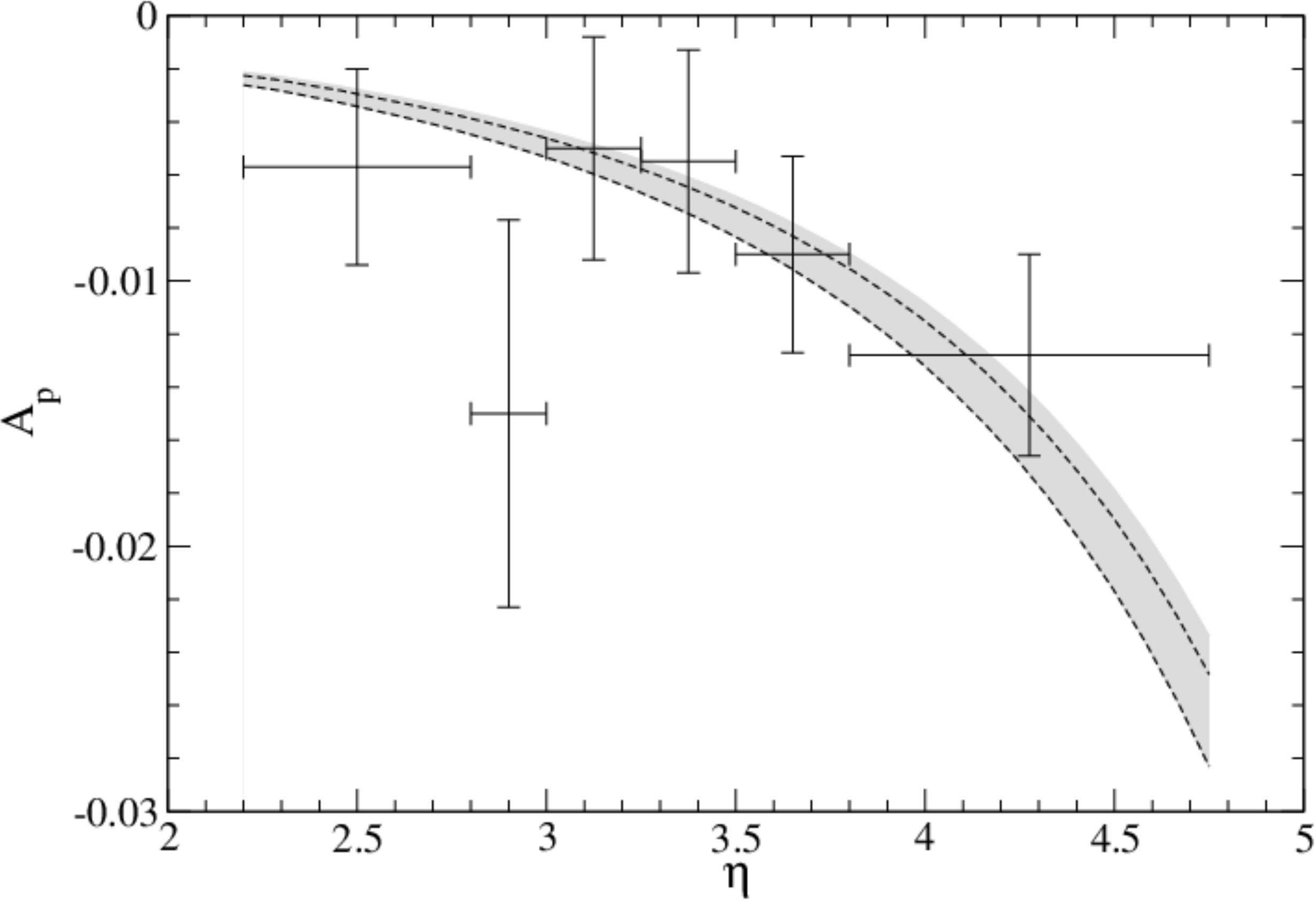}
}
\hskip .1in
\subfigure[]{
      \includegraphics[width=0.45\textwidth,angle=0,clip]{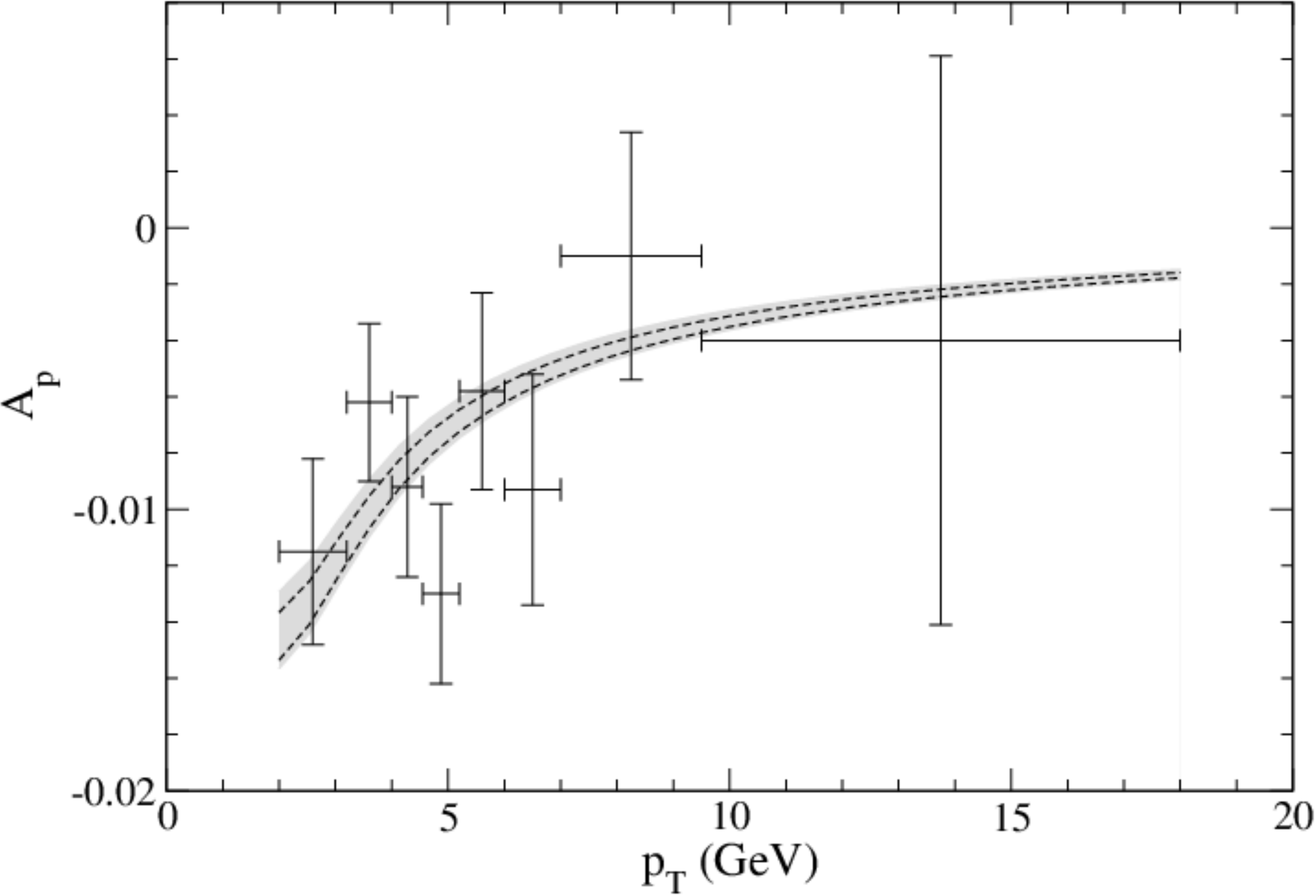}
}

\vskip .1in  
\caption[]{Asymmetry in $D^\pm$ production $A_p$ as a function of (a) pseudorapidity $\eta$ and (b) transverse momentum $p_T$ in $7$~TeV $pp$ collisions. The data points are from LHCb~\cite{LHCb:2012fb}. The grey band is obtained by varying the $\rho$'s in the intervals
$0.055<\rho^{sm}_1<0.065$, $0.65<\rho^{sm}_8<0.8$, $0.24<\rho^{sf}_1<0.30$ and $0.24<\rho^{sf}_8<0.30$ respectively. The dashed lines are from varying $0.055 < \epsilon_c < 0.069$. }
\label{fg:Ap_dis}
\end{center}
\end{figure}

\begin{figure}[H]
\begin{center}
\subfigure[]{
      \includegraphics[width=0.45\textwidth,angle=0,clip]{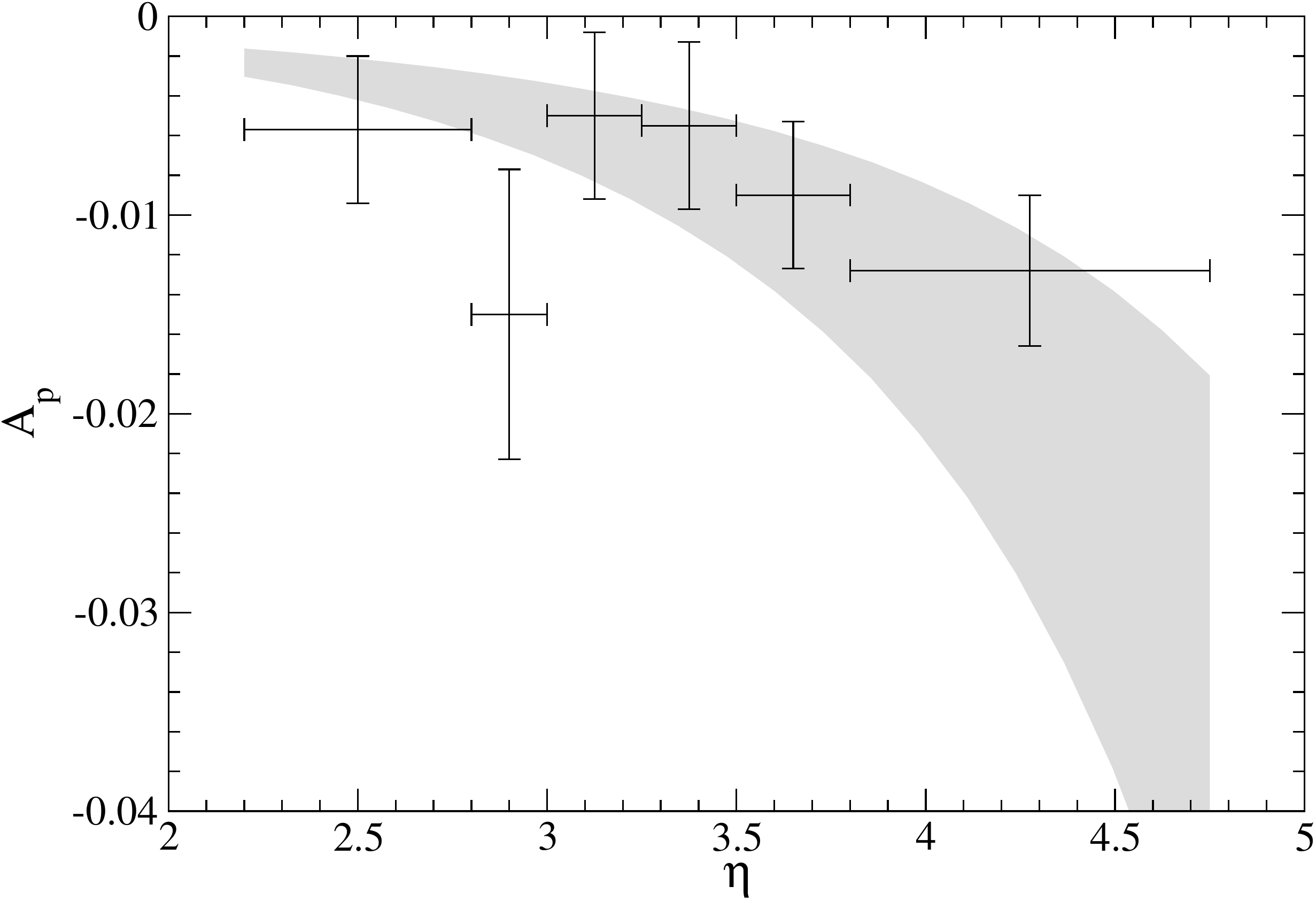}
}
\hskip .1in
\subfigure[]{
      \includegraphics[width=0.45\textwidth,angle=0,clip]{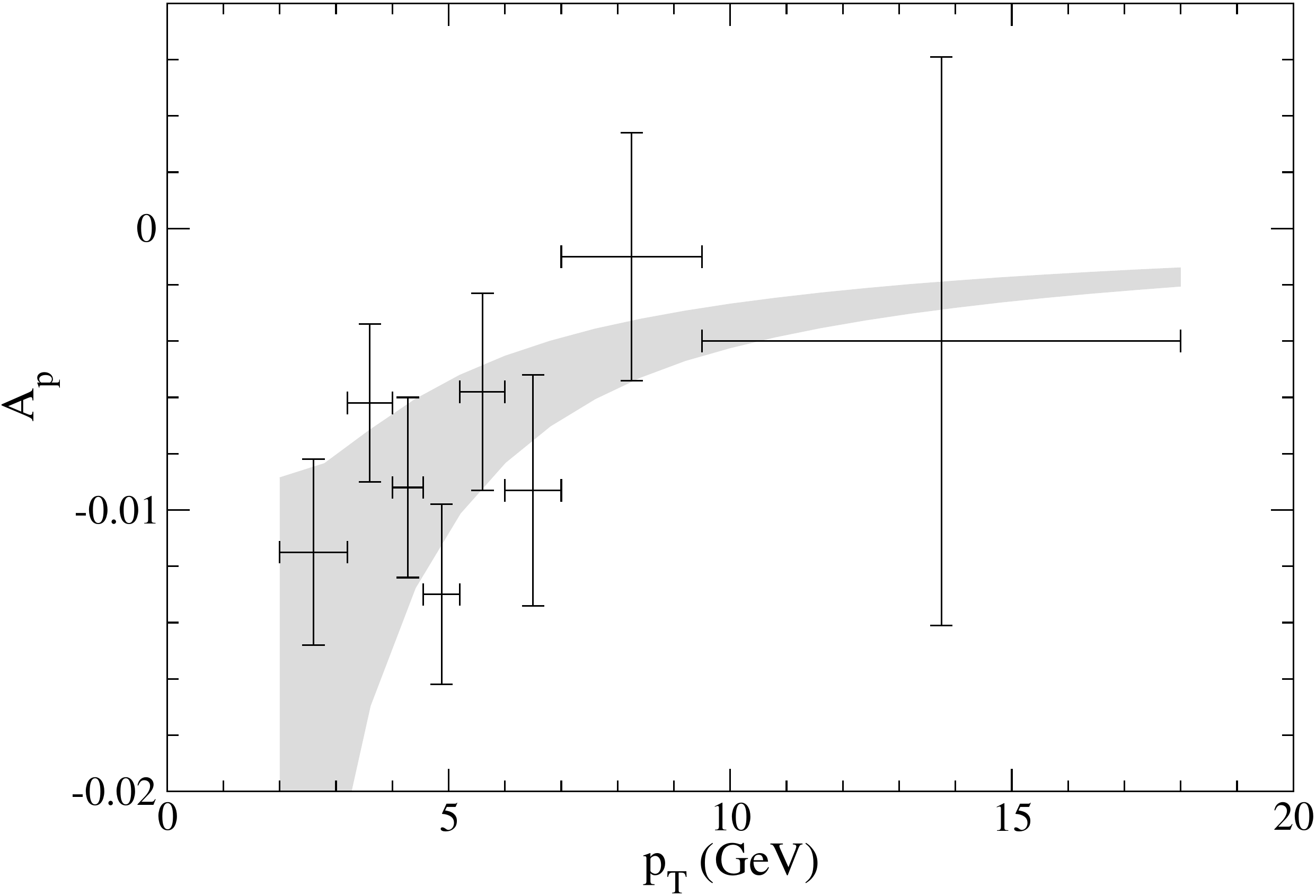}
}

\vskip .1in  
\caption[]{Asymmetry in $D^\pm$ production $A_p$ as a function of (a) pseudorapidity $\eta$ and (b) transverse momentum $p_T$ in $7$~TeV $pp$ collisions. The data points are from LHCb~\cite{LHCb:2012fb}. The grey band is obtained by varying $\mu_f$ in the interval 
$\frac{1}{2}\sqrt{p_T^2+m_c^2}<\mu_f<2\sqrt{p_T^2+m_c^2}$. The $\rho$'s are taken to be central values of the ranges in Fig.~\ref{fg:Ap_dis}.
$\epsilon_c$ is taken to be $0.06$.}
\label{fg:Ap_dis_scale}
\end{center}
\end{figure}

When integrated over $2\textrm{ GeV}<p_T<18\textrm{ GeV}$ and $2.2<\eta<4.75$, excluding the region with
$2\textrm{ GeV}<p_T<3.2\textrm{ GeV}$, $2.2<\eta<2.8$, the asymmetry $A_p$ for $D^\pm$ is found to be $-0.88\%<A_p<-1.07\%$ with $0.055<\rho^{sm}_1<0.065$, $0.65<\rho^{sm}_8<0.8$, $0.24<\rho^{sf}_1<0.30$ and $0.24<\rho^{sf}_8<0.30$. Figure~\ref{fg:Ap_dis} shows $A_p$
as a function of pseudorapidity $\eta$ and transverse momentum $p_T$ of the $D^\pm$ mesons as predicted by the heavy quark recombination mechanism. Data from Ref.~\cite{LHCb:2012fb} are shown as well.  The grey band is from varying the $\rho$ parameters within the ranges above.  The dashed line is obtained using the central value of the $\rho$ parameters and varying $\epsilon_c$ within its error bars.  The calculated distributions are reasonably consistent with the data. Figure~\ref{fg:Ap_dis_scale} shows the independence of $A_p$ on the factorization scale $\mu_f$ when $\mu_f$ is varied the the range $\frac{1}{2}\sqrt{p_T^2+m_c^2}<\mu_f<2\sqrt{p_T^2+m_c^2}$. The scale dependence is significant ($\sim 100\%$) at the high $\eta$ and low $p_T$ ends. An NLO calculation for the heavy quark recombination mechanism in the future will presumably reduce this theoretical uncertainty.

\section{$\Lambda_c^{+}/\Lambda_c^{-}$ and $\Lambda_b^0/\overline{\Lambda}_b^0$
 production asymmetry from heavy quark recombination mechanism}\label{Lambda_asymmetry}
Similarly, one can apply the heavy quark recombination mechanism to predict production asymmetries for baryons. The contributions to
production of $\Lambda_Q$ $(udQ)$ from recombination are given by
\begin{subequations}
\begin{align}
(a)\phantom{aaaa}d\hat\sigma[\Lambda_Q]&=d\hat{\sigma}[qg\rightarrow (Qq)^n+\overline{Q}]\eta[(Qq)^n\rightarrow \Lambda_Q]\,,\label{eq:recombination_a} \\
(b)\phantom{aaaa}d\hat\sigma[\Lambda_Q]&=d\hat{\sigma}[Qq\rightarrow (Qq)^n+g]\eta[(Qq)^n\rightarrow \Lambda_Q]\,,
\label{eq:recombination_b}\\
(c)\phantom{aaaa}d\hat\sigma[\Lambda_Q]&=\sum\limits_nd\hat{\sigma}[qg\rightarrow (\overline{Q}q)^n+Q]\sum\limits_{\overline{H}_{meson}}\rho[(\overline{Q}q)^n\rightarrow \overline{H}_{meson}]\otimes
D_{Q\rightarrow \Lambda_Q}\,, \label{eq:opp_recombination_meson}\\
(d)\phantom{aaaa}d\hat\sigma[\Lambda_Q]&=\sum\limits_nd\hat{\sigma}[\bar{q}g\rightarrow (\overline{Q}\bar{q})^n+Q]\sum\limits_{\overline{H}_{baryon}}\eta[(\overline{Q}\bar{q})^n\rightarrow \overline{H}_{baryon}]\otimes
D_{Q\rightarrow \Lambda_Q}\,.
\label{eq:opp_recombination_baryon}
\end{align}
\end{subequations}
Here we include both mesons and baryons in the opposite-side recombination (c) and (d). For simplicity, we will take $H$ to be a low-lying heavy hadron. Thus, for $\Lambda_c$ production we will take $H_{meson}$ to be either $D$ or $D^*$, and $H_{baryon}$ be any baryon from the lowest mass $J^p=\frac{1}{2}^+$ and $\frac{3}{2}^+$ heavy baryon $SU(3)$ flavor multiplets, and similarly for $\Lambda_b$ production. We will also assume $SU(3)$ flavor symmetry. 
All the $\rho$'s and $\eta$'s scale as $\Lambda_{QCD}/m_Q$. The leading nonperturbative parameters $\eta[(Qq)^n\rightarrow \Lambda_Q]$ are
\begin{align}
\eta_3&=\eta[Qq(^1S_0^{(\bar{3})})\rightarrow \Lambda_Q]\,,
&\tilde{\eta}_3&=\eta[Qq(^3S_1^{(\bar{3})})\rightarrow \Lambda_Q]\,, \nonumber \\
\eta_6&=\eta[Qq(^1S_0^{(6)})\rightarrow \Lambda_Q]\,,  
&\tilde{\eta}_6&=\eta[Qq(^3S_1^{(6)})\rightarrow \Lambda_Q]\,. \label{eq:rho_Lambda}
\end{align}
Contributions of feeddown from heavier baryons in processes (a) and (b) can be taken into account by using the inclusive parameter $\eta_{inc}$:
\begin{equation}
\eta_{inc}[(Qq)^n\rightarrow\Lambda_Q]=\eta[(Qq)^n\rightarrow\Lambda_Q]+\sum\limits_{H_{baryon}\neq\Lambda_Q}\eta[(Qq)^n\rightarrow H_{baryon}]
B[H_{baryon}\rightarrow\Lambda_Q+X] \label{eq:eta_inc}
\end{equation}
Here again we assume $H_{baryon}$ is a member of the lowest mass $J^p=\frac{1}{2}^+$ and $\frac{3}{2}^+$ heavy baryon $SU(3)$ flavor multiplets.

\begin{figure}[]
\begin{center}
\subfigure[]{
      \includegraphics[width=0.45\textwidth,angle=0,clip]{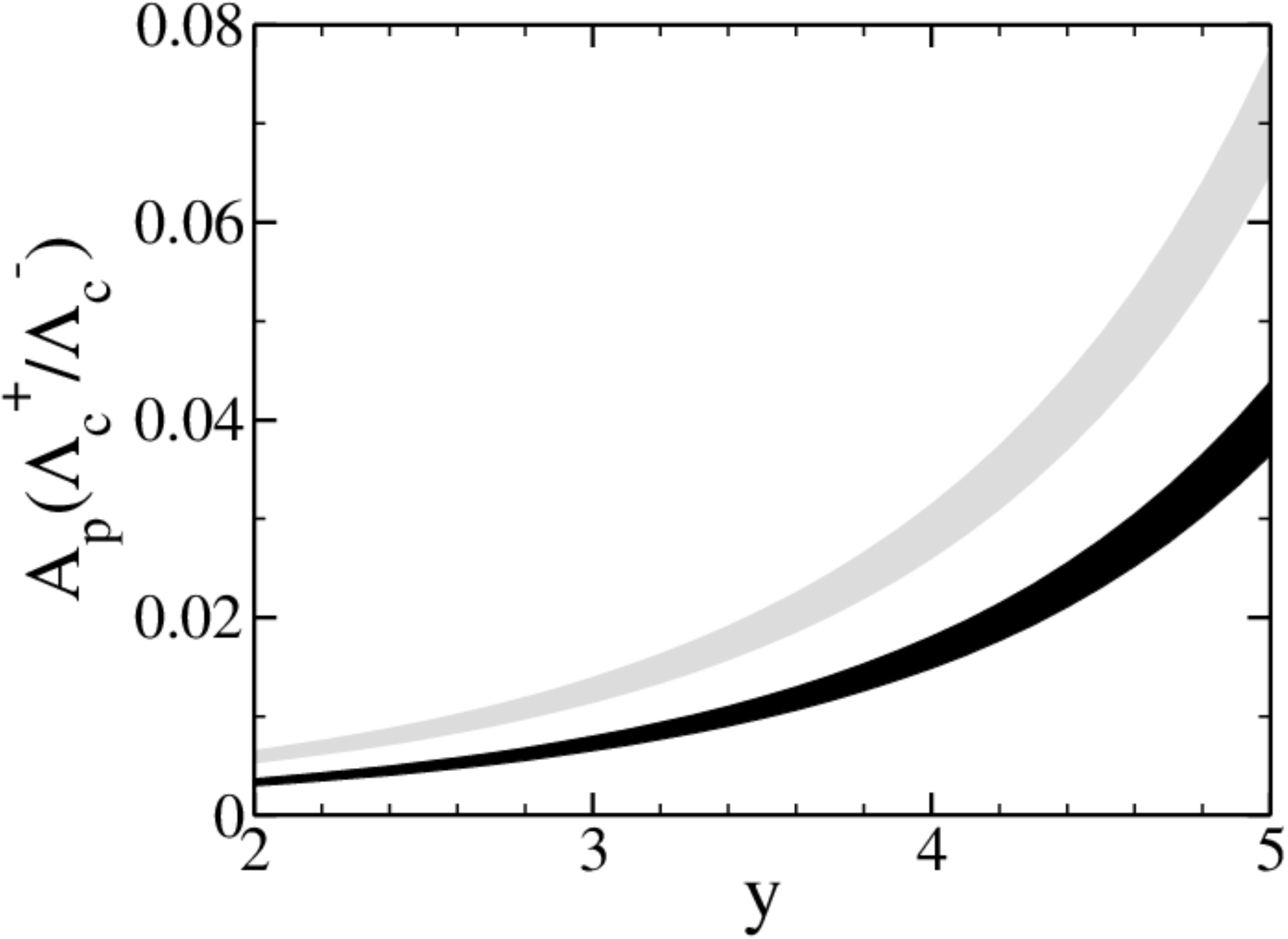}
}
\hskip .001in
\subfigure[]{
      \includegraphics[width=0.45\textwidth,angle=0,clip]{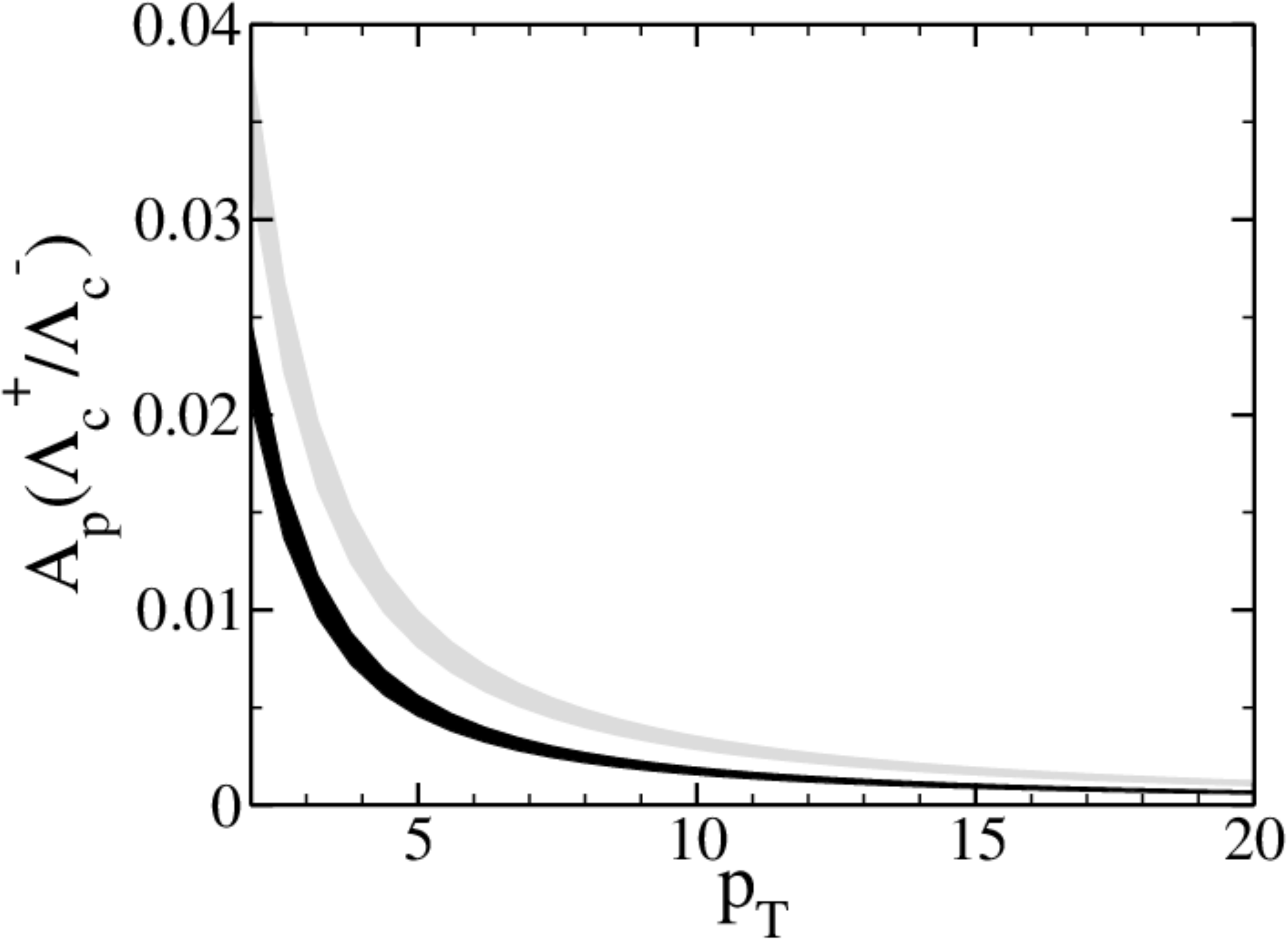}
}

\vskip .1in  
\caption[]{Asymmetry in $\Lambda_c^{+}/\Lambda_c^{-}$ production as a function of (a) rapidity $y$ and (b) transverse momentum $p_T$ in the kinematic region $2<y<5$ and $2{\rm{~GeV}}<p_T<20{\rm{~GeV}}$ in $7$~TeV (grey band) and $14$~TeV (black band) $pp$ collisions. The integrated $A_p$ is found to be $2.0\%<A_p(\Lambda_c^{+}/\Lambda_c^{-})<2.4\%$ for $\sqrt{s}=7$~TeV and $1.2\%<A_p(\Lambda_c^{+}/\Lambda_c^{-})<1.5\%$ for $\sqrt{s}=14$~TeV.}
\label{fg:Ap_c}
\end{center}

\vskip 0.5in  

\begin{center}
\subfigure[]{
      \includegraphics[width=0.45\textwidth,angle=0,clip]{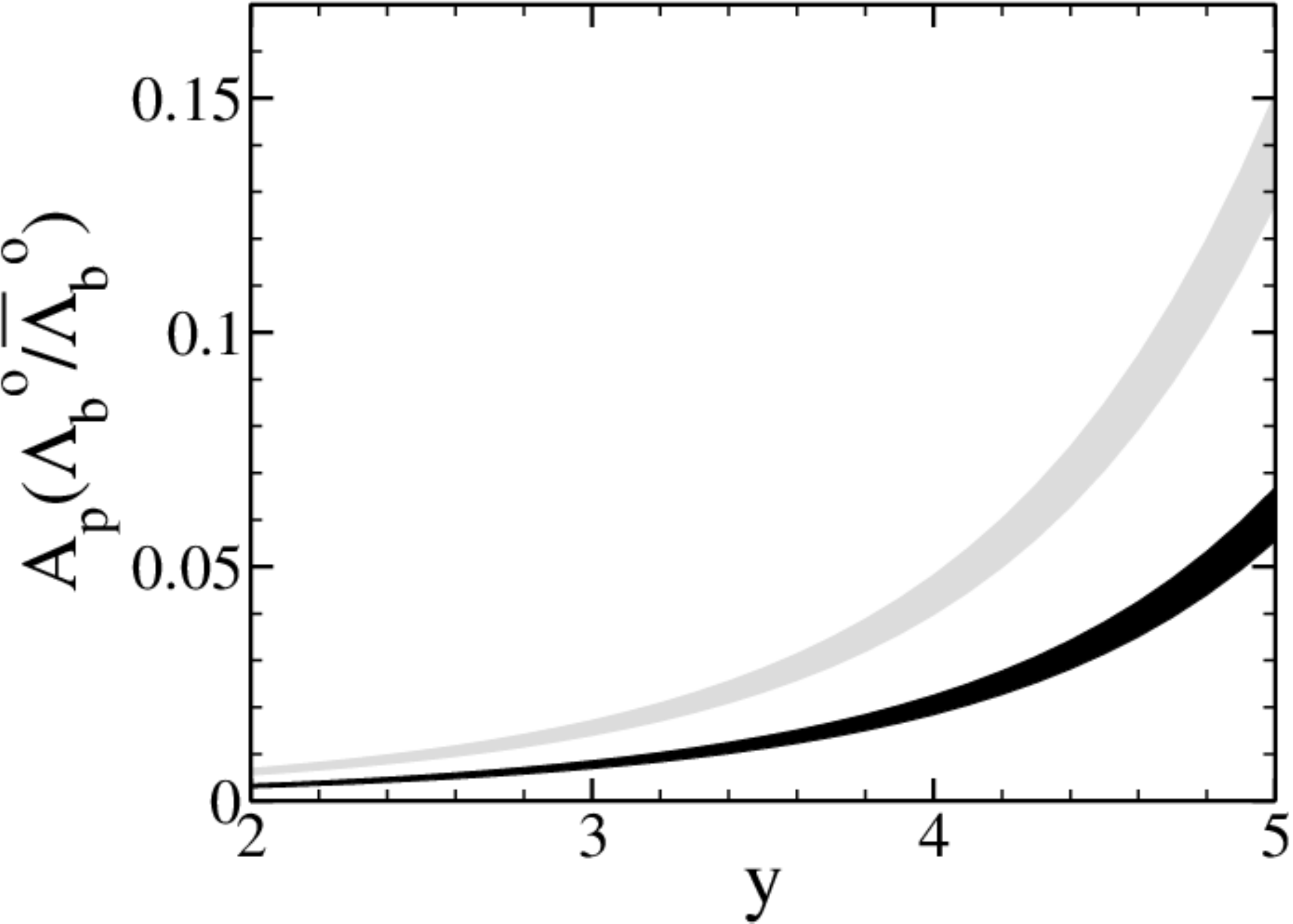}
}
\hskip .001in
\subfigure[]{
      \includegraphics[width=0.45\textwidth,angle=0,clip]{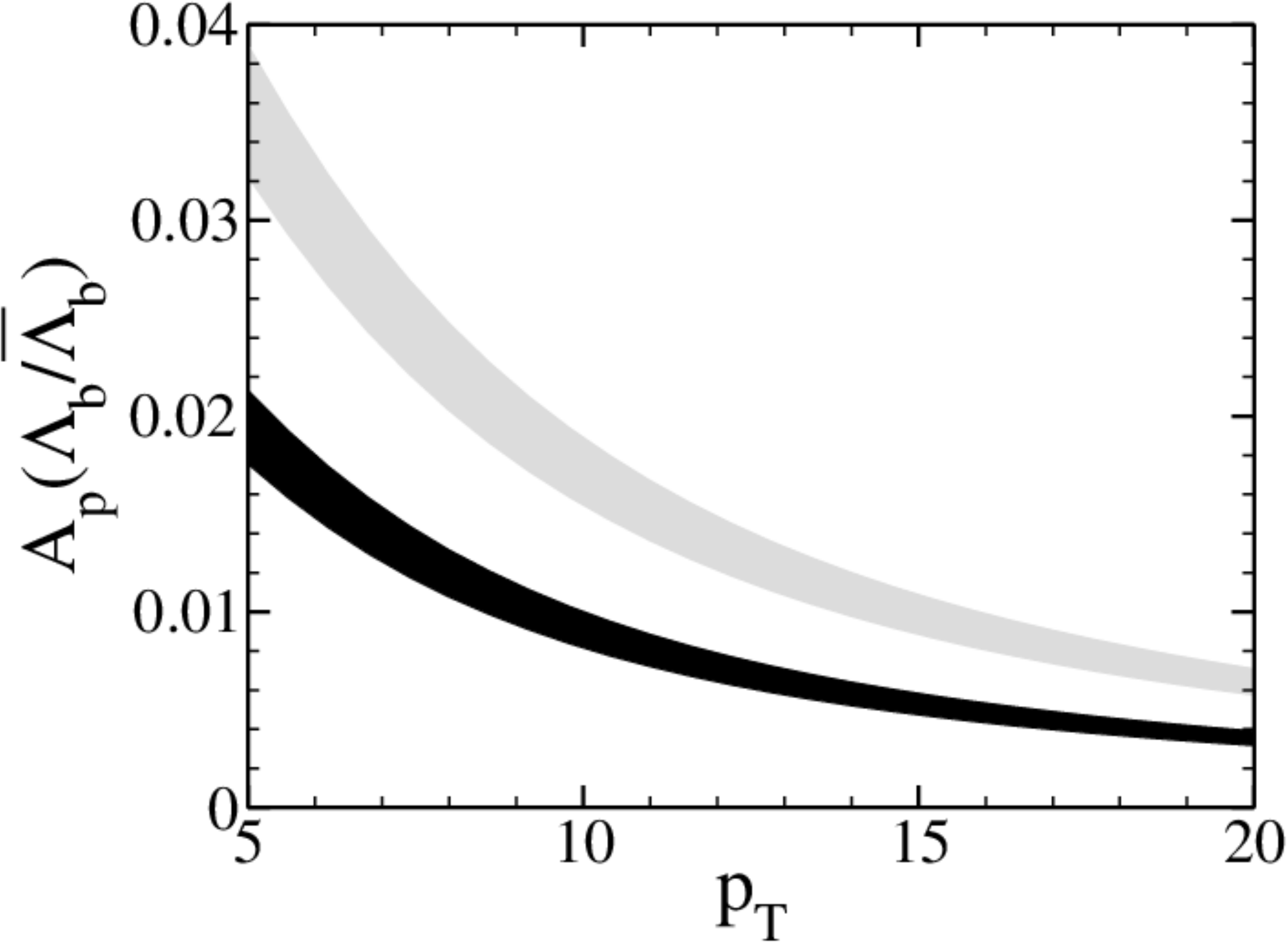}
}

\vskip .1in  
\caption[]{Asymmetry in $\Lambda_b^0/\overline{\Lambda}_b^0$ production as a function of (a) rapidity $y$ and (b) transverse momentum $p_T$ in the kinematic region $2<y<5$ and $5{\rm{~GeV}}<p_T<20{\rm{~GeV}}$ in $7$~TeV (grey band) and $14$~TeV (black band) $pp$ collisions.
The integrated $A_p$ is found to be
$2.2\%<A_p(\Lambda_b^0/\overline{\Lambda}_b^0)<2.6\%$ for $\sqrt{s}=7$~TeV and $1.1\%<A_p(\Lambda_b^0/\overline{\Lambda}_b^0)<1.4\%$ for $\sqrt{s}=14$~TeV.}
\label{fg:Ap_b}
\end{center}
\end{figure}

For the four $\rho$'s for $D$, we use the ranges in Section~\ref{D_asymmetry}. Best single-parameter fit to $\Lambda_c^{\pm}$ asymmetry in fixed target experiments gives $\tilde{\eta}_{3,inc}=0.058$ for $\Lambda_c$~\cite{Braaten:2003vy}. We will take $\eta_{3,inc}=\eta_{6,inc}=\tilde{\eta}_{6,inc}=0$ and
$0.052<\tilde{\eta}_{3,inc}<0.064$ for $\Lambda_c$. For the $\eta$s for $\Lambda_b$ and $\rho$s for $B$, we simply multiply the $\Lambda_c$ and $D$ counterparts by the theoretical scaling factor $m_c/m_b$. 
We use MSTW 2008 LO central PDFs with $m_c = 1.275$~GeV and $m_b = 4.18$ GeV. The fragmentation function $D_{Q\rightarrow \Lambda_Q}$ is taken as
\begin{equation}
D_{Q\rightarrow \Lambda_Q}(z)=f_{\Lambda_Q}\delta(1-z)\,,  \label{eq:Peterson}
\end{equation}
where $f_{\Lambda_Q}$ is the inclusive fragmentation probability. This form of fragmentation function was found to be better than the Peterson form when fitting to fixed target $\Lambda_c^{+}/\Lambda_c^{-}$ asymmetry data~\cite{Braaten:2003vy}. We take $f_{\Lambda^+_c}=0.101$, which is the average of the values listed in~\cite{Abramowicz:2013eja}. $f_{\Lambda^0_b}$ is taken to be $0.09$ from~\cite{Affolder:1999iq}. The factorization scale is set to be $\mu_f=\sqrt{p_T^2+m_Q^2}$. Figures \ref{fg:Ap_c} and \ref{fg:Ap_b} show the rapidity and transverse momentum distributions of $A_p$ for $\Lambda_c^{+}/\Lambda_c^{-}$ and $\Lambda_b^0/\overline{\Lambda}_b^0$ in $pp$ collisions at $7$~TeV and $14$~TeV respectively in the forward region. The asymmetry is significant at the high-rapidity and low-$p_T$ ends ($\sim 2-15\%$). 

\begin{figure}[]
\begin{center}
\subfigure[]{
      \includegraphics[width=0.45\textwidth,angle=0,clip]{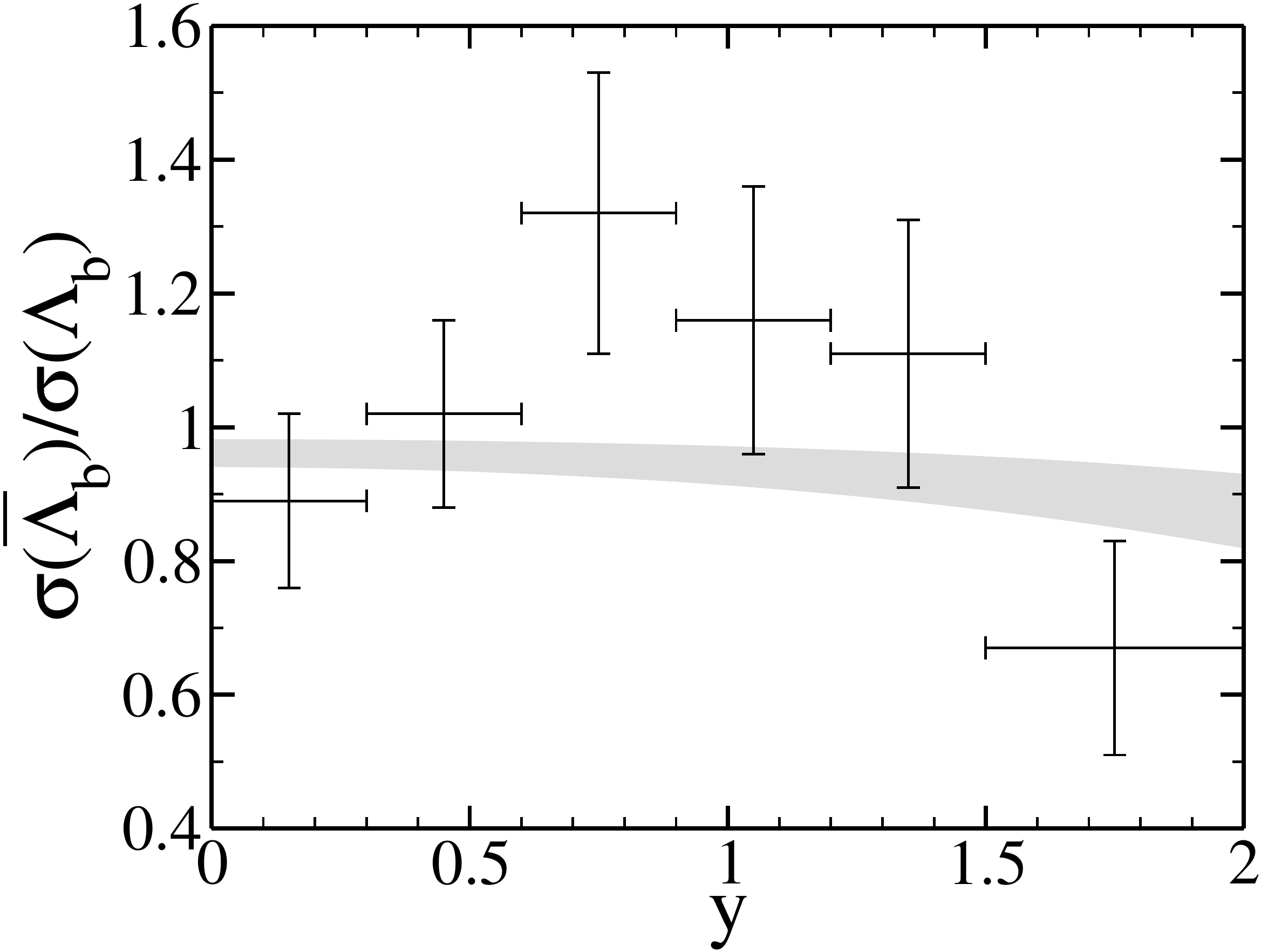}
}
\hskip .001in
\subfigure[]{
      \includegraphics[width=0.45\textwidth,angle=0,clip]{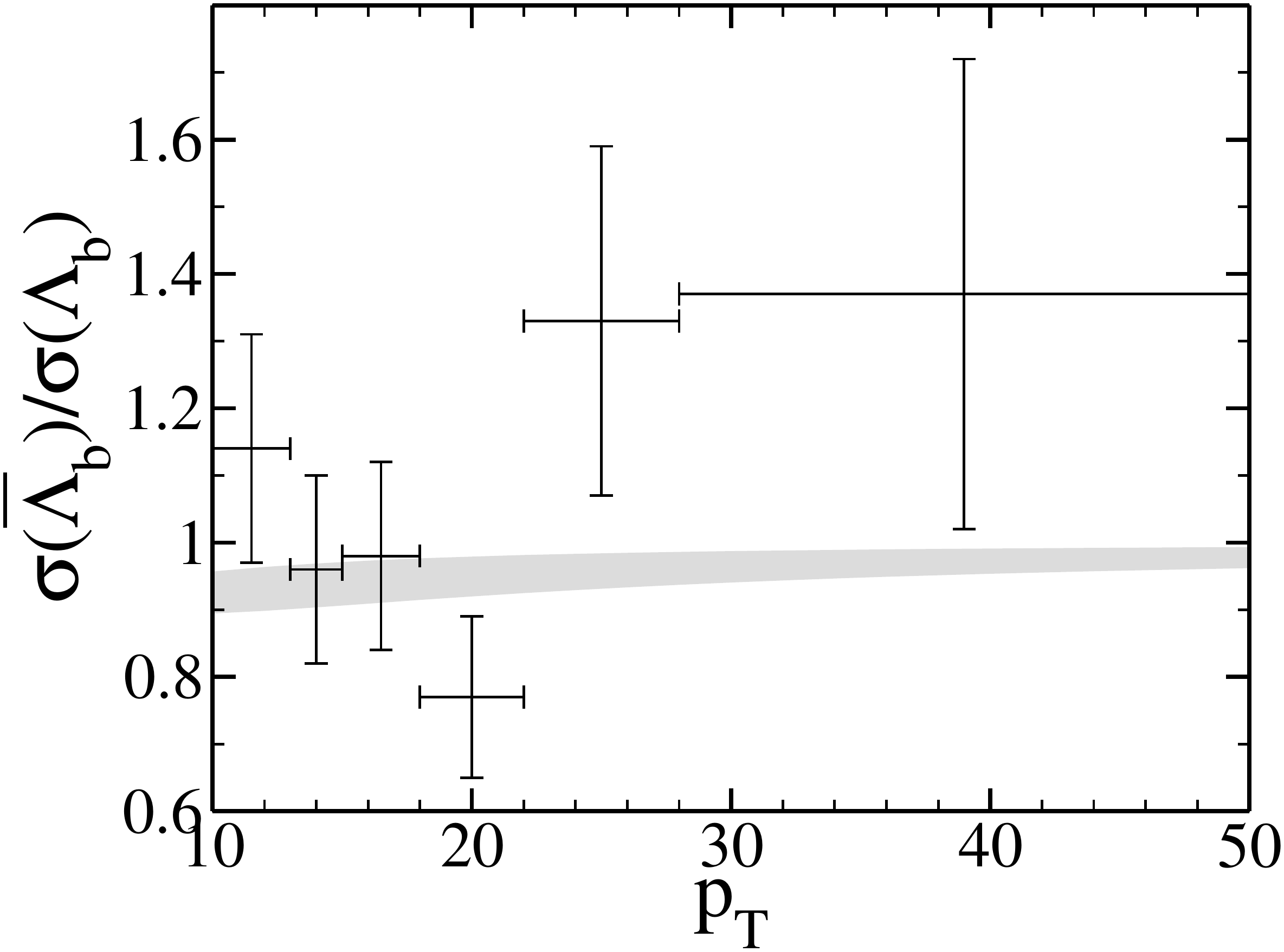}
}

\vskip .1in  
\caption[]{$\sigma(\overline{\Lambda}_b^0)/\sigma(\Lambda_b^0)$ as a function of (a) rapidity $y$ and (b) transverse momentum $p_T$ in the kinematic region $0<y<2$ and $10{\rm{~GeV}}<p_T<50{\rm{~GeV}}$ for $7$~TeV $pp$ collisions. The data are from CMS \cite{Chatrchyan:2012xg}. The grey band is our prediction from the heavy quark recombination mechanism with all $\eta_{inc}$'s set equal to each other, the range being $0.2<\eta_{inc}<1$. The ranges of $\rho$'s are as those used in Fig. \ref{fg:Ap_b}.}
\label{fg:Ap_b_CMS}
\end{center}
\end{figure}

Shown in Fig.~\ref{fg:Ap_b_CMS} are the CMS data~\cite{Chatrchyan:2012xg}, which are the rapidity and transverse momentum distributions of $\sigma(\overline{\Lambda}_b^0)/\sigma(\Lambda_b^0)$ for $7$~TeV $pp$ collisions in the kinematic region $0<y<2$ and $10{\rm{~GeV}}<p_T<50{\rm{~GeV}}$. The CMS data, despite of the large error bars, do have a slight trend of surplus of $\Lambda_b^0$ over $\overline{\Lambda}_b^0$ in the regions of high rapidity and low transverse momentum respectively. However, with the values of $\eta$'s we used above, the asymmetry predicted from the heavy quark recombination mechanism is negligible in this kinematic region. To see the limit of the heavy quark recombination mechanism in explaining the data, in Fig.~\ref{fg:Ap_b_CMS} we also plot the prediction with larger values of $\eta$'s. Here all $\eta_{inc}$'s are set equal to each other, with the range being $\Lambda_{QCD}/m_b\sim0.2<\eta_{inc}<1$. The ranges of the $\rho$s are as those used in Fig.~\ref{fg:Ap_b}. Although the prediction shows a significant asymmetry $\sim10\%$ at the high-rapidity and low-$p_T$ ends, it still fails to hit the bin with the largest rapidity. We believe that our previous predictions with  smaller $\eta_{inc}$'s (Figs.~\ref{fg:Ap_c}-\ref{fg:Ap_b}) are more reasonable since those values of $\eta$'s were obtained from better fit to fixed-target experiments. Moreover, the surplus of $\overline{\Lambda}_b^0$ over $\Lambda_b^0$ at $y\sim 1$ and $p_T>20$~GeV in the CMS data cannot be explained by any existing model. We hope that data from LHCb in the future will settle the issue.
 


\section{Conclusion}\label{conclusion}

In summary, we have calculated the $D^\pm$ asymmetry using the heavy quark recombination mechanism for production at the LHCb experiment.  The measured asymmetry of $A_p=-0.96\pm0.26\pm0.18\%$ in the kinematic range $2.0\textrm{ GeV}<p_T<18\textrm{ GeV}$ and $2.2<\eta<4.75$ \cite{LHCb:2012fb} can be reproduced using reasonably sized nonperturbative parameters $\rho_{1,8}^{sm, sf}$.  Further, the $p_T$ and $\eta$ distributions are simultaneously reproduced by the heavy quark recombination mechanism.

We have also used the heavy quark recombination mechanism to calculate the production asymmetries for $\Lambda_c^{+}/\Lambda_c^{-}$ and $\Lambda_b^0/\overline{\Lambda}_b^0$ at the LHCb experiment. The differential distributions are significant at the high-rapidity and low-$p_T$ ends ($\sim 2-15\%$). The integrated asymmetries in the LHCb region are of the order of $\sim1-3\%$ and should be measurable. 

\Acknowledgements
W. K. Lai is supported in part by the National Science Foundation under Grant No. PHY-1212635. We thank A. K. Leibovich and A. A. Petrov for helpful advice and collaboration.


\end{document}